\documentclass[aps,prd,reprint,superscriptaddress]{revtex4-2}
\usepackage{amsmath,amssymb,multirow,bm}
\usepackage{soul}
\usepackage{epsfig}
\usepackage{graphicx}
\usepackage{lipsum}
\usepackage{color}
\makeatother

\usepackage{babel}
\usepackage[colorlinks,linkcolor=blue,anchorcolor=blue,citecolor=blue,urlcolor=blue,filecolor=black]{hyperref}

\newcommand{\beq}{\begin{equation}}
\newcommand{\eeq}{\end{equation}}
\newcommand{\bea}{\begin{eqnarray}}
\newcommand{\eea}{\end{eqnarray}}
\preprint{}

\begin{document}
\title{A Bayesian Neural Network for the Neutron Star Equation of State}

\author{J.D. Baker}
\affiliation{Department of Physics and Astronomy, East Texas A\&M University, Commerce, TX 75428, USA}

\author{C.A. Bertulani}
\affiliation{Department of Physics and Astronomy, East Texas A\&M University, Commerce, TX 75428, USA}

\author{R.V. Lobato}
\affiliation{Centro Brasileiro de Pesquisas Físicas, Rua Dr. Xavier Sigaud, 150, Rio de Janeiro, 22290-180, RJ, Brazil}

\date{\today}

\begin{abstract}
We present a physics-informed Bayesian neural-network framework for inferring neutron-star equations of state from theoretical priors and propagating the resulting uncertainty to stellar observables. Trained on a representative set of hadronic EoSs, the model learns the equation of state through stochastic variational inference by representing the squared speed of sound with a bounded network output and obtaining the pressure by integration, so that causality, thermodynamic stability, and monotonicity are guaranteed by construction, with low-density nuclear and perturbative-QCD normalization anchors. Core EoSs are matched to an SLy4 crust and propagated through a unified Tolman-Oppenheimer-Volkoff-plus-tidal solver to obtain posterior predictions in the mass-radius ($M$-$R$) and mass-tidal-deformability ($M$-$\Lambda$) planes. The physics-informed prior is then updated with current multi-messenger data: NICER radius measurements, the GW170817 tidal-deformability constraint, and the $2\,M_\odot$ maximum-mass bound; included directly in the variational objective. The observational update moves the canonical radius from $R_{1.4}=13.41\,\mathrm{km}$ in the prior to $R_{1.4}=12.74^{+0.97}_{-0.73}\,\mathrm{km}$ (nominal 90\% variational CI), with $\Lambda_{1.4}=428^{+249}_{-130}$ and $M_{\mathrm{max}}\gtrsim 2.0\,M_\odot$. This framework provides a non-parametric route from microphysical EoS uncertainty to neutron-star observables.
\end{abstract}

\maketitle

\section{Introduction}

The dense-matter equation of state (EoS) remains uncertain above nuclear saturation density, where controlled nuclear-theory constraints and perturbative-QCD limits leave a broad intermediate regime. With current multi-messenger measurements, the central task is no longer only to find a single acceptable EoS, but to quantify how microscopic uncertainty propagates to macroscopic observables such as mass, radius, and tidal deformability \cite{oertel/2017, komoltsev/2024, abbott/2017a, abbott/2018,riley/2019, riley/2021, choudhury/2024}.

The modern multi-messenger program was catalyzed by GW170817 and subsequent tidal-deformability analyses, which demonstrated that merger signals can directly constrain the pressure support of matter near and above saturation density \cite{abbott/2017a, abbott/2018, annala/2018}. In parallel, Bayesian inference frameworks that combine pulsar masses (massive pulsars \cite{antoniadis/2013, cromartie/2019, fonseca/2021}), X-ray pulse-profile modeling, and gravitational-wave likelihoods have become a standard route to posterior EoS constraints while making prior assumptions explicit \cite{steiner/2010, landry/2020, raaijmakers/2021}. More broadly, recent work combines machine-learning surrogates and emulators for neutron-star structure with Bayesian and semiparametric multimessenger EoS inference, including covariant-density-functional, likelihood-systematics, and speed-of-sound-constrained analyses \cite{zhou/2024, papigkiotis/2025, liodis/2024, fujimoto/2024, stergioulas/2024a, ng/2025, li/2025, li/2025, dong/2025, li/2025a, li/2025b, papigkiotis/2026}.

Neural networks have by now been applied to essentially every stage of this inference chain. Deterministic networks have been trained to map mass-radius (or spectral) data to EoS parameters or pressure values \cite{fujimoto/2018, fujimoto/2020, fujimoto/2021, morawski/2020, ferreira/2021, krastev/2022, ferreira/2022, krastev/2023, thete/2023}, to reconstruct the EoS via automatic differentiation through the stellar-structure equations \cite{soma/2022, soma/2023}, and to classify EoS families from observables \cite{goncalves/2023}. Nonparametric neural representations of the EoS itself have been sampled within fully Bayesian multimessenger analyses \cite{han/2021, zhou/2023}, Bayesian neural networks have been used to decode neutron-star composition \cite{carvalho/2023}, and uncertainty-aware and simulation-based-inference pipelines have been developed to extract EoS information directly from telescope spectra \cite{farrell/2023, farrell/2023a, brandes/2024, ventagli/2025}. Relative to this body of work, the present framework combines two inference stages: it first constructs a physics-informed prior over EoS function space from a pooled ensemble of tabulated hadronic models, and then updates the network-weight posterior by placing a multi-messenger likelihood directly inside the variational objective. The stellar observations therefore inform the inferred EoS itself.

The physics-informed prior is built from hadronic EoSs in the CompOSE database \cite{typel/2022}. Rather than regressing pressure directly, the Bayesian network outputs a bounded squared speed of sound, $c_s^2(\epsilon)\in(0,1)$, and pressure is obtained by integration. Causality, thermodynamic stability, positivity, and monotonicity are therefore structural properties of every core-EoS draw. Three low-density nuclear anchors and a broad perturbative-QCD anchor set the normalization of the learned functions. Each posterior core EoS is joined to an SLy4 crust and propagated through the exact TOV-plus-tidal solver \cite{tolman/1939, oppenheimer/1939, douchin/2001}.

To perform the observational update, we use a differentiable emulator of the EoS-to-observable map inside Stochastic Variational Inference (SVI). The likelihood combines the NICER mass-radius measurements of PSR J0030+0451, PSR J0740+6620, and PSR J0437-4715 \cite{riley/2019, riley/2021, fonseca/2021, choudhury/2024, reardon/2024}, the GW170817 tidal-deformability constraint \cite{abbott/2018}, and the observed $2\,M_\odot$ maximum-mass bound \cite{antoniadis/2013, cromartie/2019, fonseca/2021}. Gradients through the emulator update the BNN weights during SVI, while all reported mass-radius and mass-tidal-deformability results are subsequently recomputed with the exact TOV-plus-tidal solver.

Unlike low-dimensional parametric families such as piecewise polytropes and spectral expansions \cite{read/2009, lindblom/2010}, the resulting model defines a posterior directly over EoS functions. Relative to our previous BPR-HE analysis \cite{chimanski/2023}, the present framework removes the fixed analytic pressure ansatz, enforces core-EoS admissibility by construction, and incorporates current multi-messenger information as an explicit likelihood rather than an a posteriori comparison. Its principal contributions are therefore the structurally admissible speed-of-sound parametrization, the differentiable observational update of the EoS posterior, and the unified propagation of the updated ensemble to both the $M$-$R$ and $M$-$\Lambda$ planes.

\subsection{The equation of state}
\label{sec:eos}

The equation of state (EoS) of nuclear matter encodes the relationship between pressure, density, and composition of strongly interacting matter. In neutron stars it determines the mass-radius relation, tidal deformabilities measured in gravitational-wave events, and the maximum mass, linking astrophysical observations to the microphysics of QCD in the non-perturbative regime \cite{oertel/2017, chatziioannou/2025}.

Despite sustained progress, inferring the EoS remains challenging because the relevant densities span several orders of magnitude, from the crust ($\rho\ll\rho_0$) to the inner core ($\rho\gtrsim 5\rho_0$), where $\rho_0$ is the saturation density. In this regime, first-principles calculations are limited: chiral effective field theory provides controlled constraints at and below a few times $\rho_0$, typically for pure neutron matter and symmetric nuclear matter rather than directly for charge-neutral, beta-equilibrated neutron-star matter, whereas at asymptotically high densities perturbative QCD yields an EoS band, leaving a broad intermediate-density region where modeling assumptions dominate \cite{oertel/2017,komoltsev/2024, chatziioannou/2025}.

From the nuclear-physics perspective, major uncertainties arise from the density dependence of the symmetry energy, many-body correlations, and the role of three-nucleon forces, which affect the stiffness of neutron-rich matter. Additional complications include the possible appearance of new degrees of freedom (hyperons, meson condensates, deconfined quarks), the treatment of finite-temperature effects in mergers and supernovae, and the need to enforce thermodynamic stability and causality in any extrapolation.

A further practical issue is that different microscopic approaches provide EoSs in tabulated form with different density grids, compositions, and matching prescriptions between the crust and the core. Community resources such as the CompOSE/COMPOSE database standardize access to large ensembles of tabulated EoSs and facilitate systematic comparisons and statistical inference \cite{typel/2022}.

 \subsubsection{Parametric Representations and Motivation for a Non-Parametric Approach}
 EoS inference has traditionally relied on low-dimensional parametric representations: piecewise polytropes \cite{read/2009}, spectral expansions of the adiabatic index \cite{lindblom/2010}, and empirical Taylor expansions around saturation in coefficients such as $K_0$, $J$, and $L$ \cite{lattimer/2013,margueron/2018}. These forms are computationally convenient, but each imposes structure that can leak into the inferred physics: segment boundaries produce derivative artifacts in the speed of sound, truncated bases can under-resolve or artificially oscillate around localized features, and near-saturation expansions lose formal control well above $\rho_{\text{sat}}$. In Ref.~\cite{chimanski/2023} we introduced a Bayesian power-regression model with heteroscedastic errors (BPR-HE), trained on 65 hadronic EoSs from the LIGO LALSuite infrastructure \cite{lalsuite}, which removed segment-boundary artifacts through a continuous power-law mean relation $P(\epsilon)=a\,\epsilon^{b}$ with a density-dependent spread. Its remaining limitation is that the mean trend is still tied to a fixed analytic form, which motivates the present PI-BNN framework as a non-parametric Bayesian successor that learns a posterior \emph{distribution over functions}. 

\section{Beyond Parametrization: Learning Distributions over Functions}
Traditional EoS parameterizations are intrinsically \textit{parametric}: they assume a fixed functional form and then infer a limited set of coefficients. This can restrict the allowed thermodynamic behavior and force epistemic uncertainty to be represented only through uncertainties on a few parameters.

In contrast, the PI-BNN defines a non-parametric distribution over dense-matter EoS functions. The network maps energy density to a latent function $g_\theta(\epsilon)$, whose sigmoid gives the bounded sound speed $c_s^2(\epsilon)$; pressure is then obtained by integrating $c_s^2$. Bayesian inference over the network weights therefore produces a posterior distribution over physically admissible pressure functions rather than a single best-fit curve.

This unified surrogate perspective offers two practical advantages. First, it enables flexible interpolation and controlled extrapolation across heterogeneous microscopic inputs (hadronic models, near-saturation constraints, and high-density QCD guidance) without hand-crafted segment boundaries. Second, it propagates uncertainty consistently from microphysics to macroscopic observables, yielding posterior predictive EoS bands and robust predictions for masses, radii, and tidal deformabilities.

\subsection{The CompOSE Database}
To train the Bayesian Neural Network (BNN) and anchor its predictions in established nuclear physics, we construct a comprehensive training dataset sourced from the CompOSE (CompStar Online Supernovæ Equations of State) database. CompOSE provides a standardized repository of dense matter EoS models derived from various theoretical frameworks, including relativistic mean-field theories and Skyrme interactions. 

The training set comprises 98 cold, beta-equilibrated hadronic tables, selected from CompOSE by composition alone. Hybrid tables and models containing deconfined quark matter are excluded, but no cut is imposed on the maximum TOV mass or on any other macroscopic observable. The ensemble therefore serves as a theory-side prior rather than a set preselected to satisfy astrophysical measurements; those measurements enter only through the observational likelihood described in Sec.~\ref{sec:reweighting}. Because the training set contains no hybrid-matter models, posterior draws should not be interpreted as spontaneously reconstructing first-order phase transitions absent from the training distribution.

The tabulated ranges are strongly heterogeneous. All tables begin at essentially vacuum densities in the outer crust ($\epsilon \lesssim 10^{-6}\,\text{MeV/fm}^3$), but their maximum tabulated energy densities range from $51$ to $7698\,\text{MeV/fm}^3$: $13$ of the $98$ tables terminate below $200\,\text{MeV/fm}^3$ and therefore constrain only the vicinity of the crust-core transition. We truncate the training domain at $\epsilon=2000\,\text{MeV/fm}^3$, safely above the central energy density of the maximum-mass configuration in our posterior ($\epsilon_c\simeq1460\,\text{MeV/fm}^3$), so no stable stellar configuration probes densities beyond the cut. Not every table reaches this density: $61$ of the $98$ tables extend to $2000\,\text{MeV/fm}^3$, and only $38$ extend to the cutoff of $3000\,\text{MeV/fm}^3$. Among the tables tabulated at $\epsilon=1500\,\text{MeV/fm}^3$, the subset that continues to $3000\,\text{MeV/fm}^3$ is systematically softer, with median pressure $588\,\text{MeV/fm}^3$, compared with $760\,\text{MeV/fm}^3$ for the subset terminating between $1500$ and $3000\,\text{MeV/fm}^3$. Retaining the sparsely populated tail to $3000\,\text{MeV/fm}^3$ would therefore skew the pooled high-density training data toward these softer models without informing any stable stellar configuration. Tables ending earlier contribute no points beyond their tabulated ranges, so model coverage of the pooled dataset decreases with density; accordingly, the high-density posterior is controlled increasingly by the physics anchors rather than by the training data.

This density-dependent coverage also clarifies how the training ensemble enters the BNN: the network is trained on the combined set of $(\epsilon,P)$ points from all selected tables, without model-identity labels, rather than on complete EoS curves treated as separate objects. The learned distribution should therefore be interpreted as an ensemble-level prior. At any given density, its predictive spread combines the dispersion among the hadronic tables that extend to that density with the epistemic uncertainty of the network weights. Individual posterior draws are consequently smooth EoS functions supported by the pooled ensemble, not reconstructions of any single microscopic model. We return to the implications of this construction for the low-mass region of the $M$-$R$ posterior in Sec.~\ref{sec:results_mr}.

\paragraph{Data Preprocessing and Balancing}:
Raw EoS tables frequently exhibit highly irregular sampling frequencies, typically concentrating a massive number of discrete data points in the low-density regime while sparsely sampling the high-density extremes. If fed directly into a neural network, this class imbalance would cause the loss function to disproportionately prioritize fitting the low-density behavior, degrading the model's accuracy and uncertainty quantification in the critical inner core.

To resolve this, we implement a targeted binning and refinement algorithm. The continuous domain of normalized energy density is divided into $N = 30$ uniform histograms. Within each bin, we randomly sample a maximum of $M = 100$ data points. This ensures a balanced, quasi-uniform distribution of training data across the entire density range, helping ensure that the BNN assigns comparable statistical weight across the regimes represented by the hadronic models.

Finally, neural-network optimization is sensitive to the scale of the input and target variables. Because energy density and pressure span several orders of magnitude, we use the global maximum values $\epsilon_{\max}=\max(\epsilon)$ and $P_{\max}=\max(P)$ to define
\begin{equation}
\tilde{\epsilon} = \frac{\epsilon}{\epsilon_{\max}}, \qquad \tilde{P} = \frac{P}{P_{\max}}.
\end{equation}
The network receives $\tilde{\epsilon}$ as input, but its sigmoid output is the dimensionless physical quantity $c_s^2$. The pressure is reconstructed by integrating this output over the corresponding physical energy-density grid and is divided by $P_{\max}$ only when compared with the normalized pressure targets. Consequently, no derivative of a normalized pressure surrogate, and no chain-rule rescaling of such a derivative, enters the present formulation. A numerical offset of $10^{-5}$ is added to normalized arrays where needed to avoid operations at exact zero. All subsequent equations are written in the physical variables $\epsilon$ and $P$.

\subsection{Physics-Informed Bayesian Neural Network}
The surrogate is a fully connected network that maps normalized energy density to the latent function $g_\theta(\epsilon)$. Its hidden layers use the smooth \texttt{Softplus} activation, which keeps the resulting $c_s^2(\epsilon)$ and integrated pressure smooth. A deterministic network would provide only one extrapolation in the data-sparse inner-core regime; we therefore assign probability distributions to the weights and biases so that uncertainty in the network parameters propagates to uncertainty in the EoS and stellar observables. Independent Gaussian weight priors and a mean-field Gaussian variational posterior are used in the present implementation. The training likelihood compares the integrated pressure $P_\theta(\epsilon_i)$ with each tabulated pressure $P_i$, while the low-density and pQCD anchors enter as additional log-density terms. The multi-messenger likelihood introduced in Sec.~\ref{sec:reweighting} subsequently updates this physics-informed prior.

\subsubsection{BNN Architecture}
\label{sec:cs2_arch}
The BNN weights and biases $\theta$ are treated as random variables, so uncertainty in the latent function propagates to both $c_s^2(\epsilon)$ and the integrated pressure $P_\theta(\epsilon)$.

Rather than regressing the pressure directly, the network outputs the squared speed of sound as a function of energy density, and the pressure is obtained by integration. Concretely, the network $g_\theta(\epsilon)$ maps energy density to a bounded squared sound speed through a sigmoid output activation,
\begin{equation}
c_s^2(\epsilon) = \mathrm{sigmoid}\!\left[g_\theta(\epsilon)\right] \in (0,1),
\end{equation}
and the pressure follows from the thermodynamic identity $c_s^2 = (\partial P/\partial\epsilon)_S$,
\begin{equation}
P(\epsilon) = P_0 + \int_{0}^{\epsilon} c_s^2(\epsilon')\,\mathrm{d}\epsilon',
\label{eq:cs2_integral}
\end{equation}
where $P_0$ is the surface pressure fixed by the crust match. This parametrization is the central methodological change relative to direct pressure regression. The sigmoid confines the sound speed to $0<c_s^2<1$, so every core draw is causal and thermodynamically stable; because pressure is the integral of a positive quantity, it is also positive and monotonically increasing. No EoS-level acceptance criterion is therefore applied: every posterior core draw is retained. The integral in Eq.~(\ref{eq:cs2_integral}) is evaluated by the trapezoidal rule on a fixed energy-density grid, so the map $\theta\mapsto P_\theta(\epsilon)$ remains differentiable for gradient-based variational inference.

We define independent Gaussian priors over all weights and biases,
\begin{equation}
p(\theta) = \prod_i \mathcal{N}(\theta_i | 0, \sigma_{w}^2),
\end{equation}
with prior scale $\sigma_{w} = 0.1$. The hidden layers use the smooth \texttt{Softplus} activation, which keeps $c_s^2(\epsilon)$ continuously differentiable and therefore yields a smooth pressure profile.

To approximate the intractable true posterior $p(\theta | \mathcal{D})$, we use Stochastic Variational Inference (SVI). We introduce a parametrized variational distribution $q_\phi(\theta)$ (a mean-field Gaussian) and optimize the variational parameters $\phi$ by maximizing the Evidence Lower Bound (ELBO):
\begin{equation}
\mathcal{L}_{\text{ELBO}}(\phi) = \mathbb{E}_{q_\phi(\theta)}[\log p(\mathcal{D} | \theta)] - D_{\text{KL}}(q_\phi(\theta) || p(\theta))
\end{equation}
The likelihood is evaluated using the integrated pressure,
\begin{equation}
P_i \sim \mathcal{N}\!\left(P_\theta(\epsilon_i),\sigma_i^2\right), \qquad
\sigma_i=\sigma_{\rm rel}P_i+\sigma_{\rm floor},
\end{equation}
with $\sigma_{\rm rel}=0.15$ and a small numerical floor. This is the same heteroscedastic likelihood specified in Sec.~III~A.

An MLP is sufficient because the learned relation is the one-dimensional map $\epsilon\mapsto c_s^2$. The smooth \texttt{Softplus} hidden activation and sigmoid output yield a continuously differentiable sound-speed function and a smooth integrated pressure. The mean-field Gaussian variational family provides a practical compromise between posterior flexibility and tractability, although its effect on uncertainty calibration is assessed explicitly in Sec.~\ref{sec:reweighting}.

\subsection{Physics-Informed Constraints}
\label{sec:soft_constraints}
A physical equation of state must respect thermodynamic stability and causality at all densities. As described in Sec.~\ref{sec:cs2_arch}, these requirements are enforced structurally by the $c_s^2$-integral parametrization: the sigmoid confines the sound speed to $0<c_s^2<1$, and integrating this positive quantity makes the pressure positive and monotonic. Consequently, no monotonicity or causality penalties and no EoS-level acceptance criterion are used; every posterior core draw is retained. The remaining physics input concerns normalization rather than admissibility: the EoS is anchored to nuclear physics at low density and to perturbative QCD (pQCD) at asymptotically high density.

In our probabilistic programming framework (\texttt{Pyro}), these anchors are implemented through \texttt{pyro.factor} statements, which add custom log-density terms directly to the Evidence Lower Bound (ELBO) during Stochastic Variational Inference, steering the variational posterior toward the physically anchored regions of function space while preserving the flexibility of the neural surrogate.

We use the following anchor terms during SVI:
\begin{enumerate}
    \item Low-density nuclear anchors: rather than a single saturation point, we anchor the low-density EoS at three energy densities, $\epsilon \in \{150, 300, 600\}\,\text{MeV/fm}^3$, with target pressures $\{2.9, 24.8, 124.1\}\,\text{MeV/fm}^3$ and Gaussian widths $\{0.5, 3.0, 15.0\}\,\text{MeV/fm}^3$, respectively. Using three anchors rather than one constrains the low-density \emph{shape}, which controls the radius of canonical-mass stars through the pressure near $1$-$2\,n_0$.
    \item Perturbative QCD (pQCD) anchor: we place the high-density anchor at $\epsilon_{\text{pQCD}} = 10{,}000\,\text{MeV/fm}^3$, deep in the pQCD-informed regime \cite{komoltsev/2022}, with a target pressure $P_{\text{pQCD}} = 2300\,\text{MeV/fm}^3$ and a broad Gaussian width $\sigma_{\text{pQCD}} = 1000\,\text{MeV/fm}^3$. The target is set \emph{below} the exact conformal value $P=\epsilon/3\simeq 3333\,\text{MeV/fm}^3$ because the perturbative pressure at these densities is sub-conformal, and the broad width encapsulates the residual renormalization-scale uncertainty. We stress that this anchor lies far above the central densities reached by any stable stellar configuration in our posterior, so its role is to guide the high-density extrapolation rather than to constrain the neutron-star-density EoS directly; the $c_s^2$ plateau it induces above $\epsilon\approx 1500\,\text{MeV/fm}^3$ is therefore a prior-controlled extrapolation, not a data-constrained result.
\end{enumerate}

The anchor target pressures are the medians of the hadronic training ensemble at the three densities, and the saturation value is independently consistent with $P(n_0)\approx L\,n_0/3$ for the empirical slope range $L\simeq 40$-$70$~MeV. The widths $\{0.5, 3.0, 15.0\}\,\text{MeV/fm}^3$ are set to approximately half the ensemble standard deviation at each anchor density ($\{1.0, 5.6, 31.5\}\,\text{MeV/fm}^3$), so that the anchors constrain the network to the central trend of the models while remaining consistent with the inter-model dispersion. The adequacy of this choice is confirmed a posteriori by the agreement of the posterior with the ensemble median at all densities (Fig.~\ref{fig:eos_posterior}).

Because causality and monotonicity are guaranteed by the $c_s^2$-integral parametrization, the anchor terms are the only physics contributions added to the ELBO, and they serve purely to normalize the EoS at the empirically controlled low-density and asymptotic high-density boundaries. The physics-informed objective is therefore an ELBO augmented only by the anchor log-densities,
\begin{equation}
\begin{aligned}
\mathcal{J}_{\text{PI}} ={}& \mathcal{L}_{\text{ELBO}} \\
&+ \; W_{\text{low}} \sum_{a} \log \mathcal{N}\bigl(P(\epsilon_a) \mid P_a, \sigma_a^2\bigr) \\
&+ \; W_{\text{pQCD}} \log \mathcal{N}\bigl(P(\epsilon_{\text{pQCD}}) \mid P_{\text{pQCD}}, \sigma_{\text{pQCD}}^2\bigr),
\end{aligned}
\end{equation}
where the sum runs over the three low-density anchors $a$ and $P(\epsilon)$ is the integrated pressure of Eq.~(\ref{eq:cs2_integral}). This implements Bayesian regularization in function space rather than parameter space, and requires no monotonicity or causality terms, since those properties are structural.

\subsubsection{Posterior Ensemble}
We draw $N=500$ posterior core EoSs $P_\theta^{(k)}(\epsilon)$ from the optimized variational distribution $q_\phi^*(\theta)$. Because the $c_s^2$-integral parametrization guarantees causality, positivity, and monotonicity for every core draw, no post-training EoS rejection or acceptance criterion is used. Every draw is stitched to the SLy4 crust and propagated through the TOV-plus-tidal solver, so the reported ensemble is not a selected subset of a larger set of variational samples.

\subsubsection{Crust-Core Stitching}

The structure of low-mass neutron stars ($M\lesssim1.0\,M_\odot$) is especially sensitive to the outer and inner crust. Although the pooled CompOSE tables extend to low density, their crust treatments are heterogeneous; we therefore use the same tabulated SLy4 EoS \cite{douchin/2001} for every realization and use the BNN draw only for the core segment above the matching interface.

Stitching does not impose physical admissibility on the core, which is already guaranteed by construction; it supplies a common low-density segment and an ordered connection to the BNN core. Let $(\epsilon_{\mathrm{crust,max}},P_{\mathrm{crust,max}})$ denote the terminal point of the SLy4 crust. For each posterior core EoS $P_\theta^{(k)}(\epsilon)$, the matched core segment contains the points satisfying
\begin{equation}
\epsilon>\epsilon_{\mathrm{crust,max}}, \qquad
P_\theta^{(k)}(\epsilon)>P_{\mathrm{crust,max}}.
\end{equation}

The final stitched tabulation concatenates the crust with this matched core segment,
\begin{equation}
\begin{aligned}
\mathcal{T}_{\mathrm{stitched}} ={}& \mathcal{T}_{\mathrm{SLy4}} \\
&\cup\Bigl\{\bigl(\epsilon,P_{\theta}^{(k)}(\epsilon)\bigr): \epsilon > \epsilon_{\mathrm{crust,max}}, \\
&\hspace{2.8em} P_{\theta}^{(k)}(\epsilon) > P_{\mathrm{crust,max}}\Bigr\},
\end{aligned}
\end{equation}
and the continuous stitched function $P(\epsilon)$ used in the TOV solver is obtained by monotone interpolation over $\mathcal{T}_{\mathrm{stitched}}$.

This procedure provides an ordered connection between the SLy4 crust and each posterior core segment. Monotone interpolation over the combined tabulation makes $P(\epsilon)$ continuous and preserves its ordering across the interface; it is not used as a substitute for monotonicity or causality penalties, which are absent from the present formulation.

We emphasize that this construction enforces continuity at the level of the thermodynamic variables but does not impose higher-order smoothness conditions (e.g., continuity of the speed of sound $c_s^2$), which is left for future refinement.

\subsection{Relativistic Structure Integration}
To map the posterior EoS samples into observable mass-radius and tidal-deformability space, we solve the Tolman-Oppenheimer-Volkoff (TOV) background equations \cite{tolman/1939, oppenheimer/1939} in geometric units ($G=c=1$):
\begin{equation}
\frac{dP}{dr} = -\frac{(\epsilon + P)(m + 4\pi r^3 P)}{r^2(1-2m/r)},
\end{equation}
\begin{equation}
\frac{dm}{dr} = 4\pi r^2 \epsilon.
\end{equation}

To ensure mathematically consistent background-structure calculations across the wide crust-to-core dynamical range, we represent the stitched EoS in log-log variables, $(\log_{10} P, \log_{10} \epsilon)$. We then use a monotonic Piecewise Cubic Hermite Interpolating Polynomial (PCHIP), which preserves the ordering of tabulated points and avoids artificial overshoot in $(\partial P / \partial \epsilon)_S$ (and therefore in $c_s^2$). This keeps the interpolated EoS thermodynamically admissible and prevents interpolation-induced superluminal artifacts during stellar-structure integration.

\subsubsection{Tidal Deformability and Love Number}
In addition to the TOV background, we solve the linear even-parity tidal perturbation equations for each stellar model using the same interpolated EoS and adaptive integrator. The unified solver evolves the background variables together with a first-order radial metric response variable $y(r)$, so that the stellar mass $M$, radius $R$, and surface tidal response are obtained in a single pass for each central condition. In compact form, the tidal-response equation is
\begin{equation}
r\frac{dy}{dr} + y^2 + y\,F(r) + r^2 Q(r) = 0,
\qquad
 y(0)=2,
\end{equation}
where $F(r)$ and $Q(r)$ are the standard background functions determined by the TOV solution and the local sound speed $c_s^2 = (\partial P / \partial \epsilon)_S$. In the same geometric units, they are given by
\begin{equation}
F(r)=\frac{1-4\pi r^2\left[\epsilon-P\right]}{1-2m/r},
\end{equation}
and
\begin{equation}
\begin{split}
Q(r)=&\frac{4\pi}{1-2m/r}
\Biggl[5\epsilon+9P
+\frac{\epsilon+P}{c_s^2}
-\frac{3}{2\pi r^2}\Biggr] \\
&-4\left[
\frac{m+4\pi r^3P}{r^2\left(1-2m/r\right)}
\right]^2.
\end{split}
\end{equation}
Integrating this equation to the stellar surface gives the surface value $y_R \equiv y(R)$ together with the compactness $C = GM/(Rc^2)$, or equivalently $C=M/R$ in geometric units.

From $y_R$ and $C$, we evaluate the quadrupolar (second-order) tidal Love number $k_2$ as
\begin{equation}
k_2 = \frac{8C^5}{5}(1-2C)^2
\frac{2+2C(y_R-1)-y_R}{\Xi(C,y_R)},
\end{equation}

\begin{equation}
\begin{aligned}
\Xi(C,y_R)= {}&2C\left[6-3y_R+3C(5y_R-8)\right]\\
&+4C^3\Bigl[13-11y_R+C(3y_R-2)\\
&\hspace{3.3em}+2C^2(1+y_R)\Bigr]\\
&+3(1-2C)^2\Bigl[2-y_R\\
&\hspace{3.3em}+2C(y_R-1)\Bigr]\ln(1-2C).
\end{aligned}
\end{equation}
The corresponding dimensionless tidal deformability is then defined by
\begin{equation}
\Lambda = \frac{2}{3} k_2 C^{-5}.
\end{equation}
Because the Bayesian posterior produces thousands of EoS realizations, the coupled TOV-plus-tidal ODE system is integrated with an explicit adaptive Runge-Kutta method, with the surface boundary condition $P(R)=0$ captured through terminal event handling. To manage the computational load of the ensemble, these integrations are executed in an embarrassingly parallel CPU framework. This unified TOV-plus-tidal integration makes the pipeline a joint inference framework for mass, radius, and tidal deformability, rather than an $M$-$R$-only mapping.

\section{Numerical Implementation}

The computational pipeline has two numerical components: probabilistic training of the Bayesian neural network (BNN) through Stochastic Variational Inference (SVI) \cite{hoffman/2013}, and integration of the coupled stellar-structure and tidal equations with explicit adaptive Runge-Kutta solvers, terminal event handling, and ensemble-parallel CPU execution. The learning stack is implemented in \texttt{PyTorch} \cite{paszke/2019} and \texttt{Pyro} \cite{bingham/2019}.

Because the BNN parameter space is high dimensional, we perform inference with SVI and approximate the weight posterior with a tractable variational family $q_\phi(\theta)$. The network represents $c_s^2(\epsilon;\theta)$ directly and generates $P_\theta(\epsilon)$ by integration. In practice, Pyro's \texttt{AutoNormal} guide provides a mean-field Gaussian approximation over the weights and biases.

The variational parameters $\phi$ are optimized by maximizing the Evidence Lower Bound (ELBO), or equivalently by minimizing the \texttt{Trace\_ELBO} loss, using the \texttt{Adam} optimizer with learning rate $0.01$ for exactly 20,000 steps. We then draw $N=500$ posterior samples for uncertainty propagation.

A key advantage of \texttt{Pyro} is that domain physics can enter the variational objective directly. The anchor terms summarized in Sec.~\ref{sec:soft_constraints}, and the observational likelihood of Sec.~\ref{sec:reweighting}, are implemented as \texttt{pyro.factor} contributions to the ELBO. In the present formulation the squared speed of sound is the direct network output rather than a derivative, so no auto-differentiation of the pressure is needed to evaluate thermodynamic admissibility; the pressure entering the anchor and likelihood terms is obtained by trapezoidal integration of $c_s^2(\epsilon)$ on a fixed grid, which keeps the full map differentiable for gradient-based inference.

For each posterior EoS draw, we solve the coupled TOV-plus-tidal system as an initial-value problem with an explicit adaptive Runge-Kutta 5(4) integrator and dynamic step sizing, starting at $r = 10^{-4}\,\text{km}$. The stitched core-crust EoS is evaluated with the monotonic log-log interpolation defined above, the radial step is capped at $\Delta r_{\text{max}} = 0.2\,\text{km}$ to control near-surface gradients, and the surface boundary condition $P(R)=0$ is captured through terminal event handling when the pressure reaches the SLy4 vacuum boundary. To manage the posterior ensemble efficiently, these stellar integrations are distributed across an embarrassingly parallel CPU pool. The resulting compactness and tidal response are then used to compute $k_2$ and $\Lambda$ for each stellar model.

\subsection{Network Architecture and Training Hyperparameters}
For reproducibility, the Bayesian surrogate is a fully connected MLP with one input node, two hidden layers of 64 neurons each, and one output node. The hidden layers use \texttt{Softplus}, and the output uses a sigmoid to produce $c_s^2$ directly. The smooth hidden activation prevents artificial structure in the sound-speed function and in the pressure obtained by integration.

All weights and biases are assigned independent Gaussian priors $\mathcal{N}(0,0.1)$. Variational training is performed with \texttt{Trace\_ELBO} Stochastic Variational Inference using the \texttt{Adam} optimizer with learning rate $0.01$ for 20,000 steps. The likelihood is heteroscedastic: each training point is assigned a Gaussian noise proportional to its pressure, $\sigma_i = \sigma_{\rm rel}\,P_i$ with $\sigma_{\rm rel}=0.15$ (plus a small floor), so that all densities are constrained at comparable relative precision rather than being dominated by the high-pressure points. The value $\sigma_{\rm rel}=0.15$ is somewhat tighter than the raw inter-model dispersion of the hadronic ensemble ($\approx 0.25$ across the core), a deliberate choice that makes the network follow the central trend of the training models closely while leaving the anchors and the observational likelihood their intended roles.

The anchor locations and target values are those given in Sec.~\ref{sec:soft_constraints}: three low-density nuclear anchors at $\epsilon \in \{150,300,600\}\,\text{MeV/fm}^3$ with targets $\{2.9,24.8,124.1\}\,\text{MeV/fm}^3$ and widths $\{0.5,3.0,15.0\}\,\text{MeV/fm}^3$, and a single pQCD anchor at $\epsilon_{\rm pQCD}=10{,}000\,\text{MeV/fm}^3$ with target $P_{\rm pQCD}=2300\,\text{MeV/fm}^3$ and width $\sigma_{\rm pQCD}=1000\,\text{MeV/fm}^3$. No monotonicity or causality penalties appear, since those properties are enforced structurally by the $c_s^2$-integral parametrization (Sec.~\ref{sec:cs2_arch}); the anchor weights therefore set only how tightly the posterior is held to the empirical low-density shape and the pQCD asymptote. The pressure entering each anchor term is the integrated pressure of Eq.~(\ref{eq:cs2_integral}), evaluated on a fixed grid of $600$ energy-density points spanning from the surface up to the pQCD anchor density by the trapezoidal rule.

\section{Results and Discussion}

\subsection{Microphysical Posterior and Thermodynamic Stability}
\begin{figure*}[t]
    \centering
    \includegraphics[width=\textwidth]{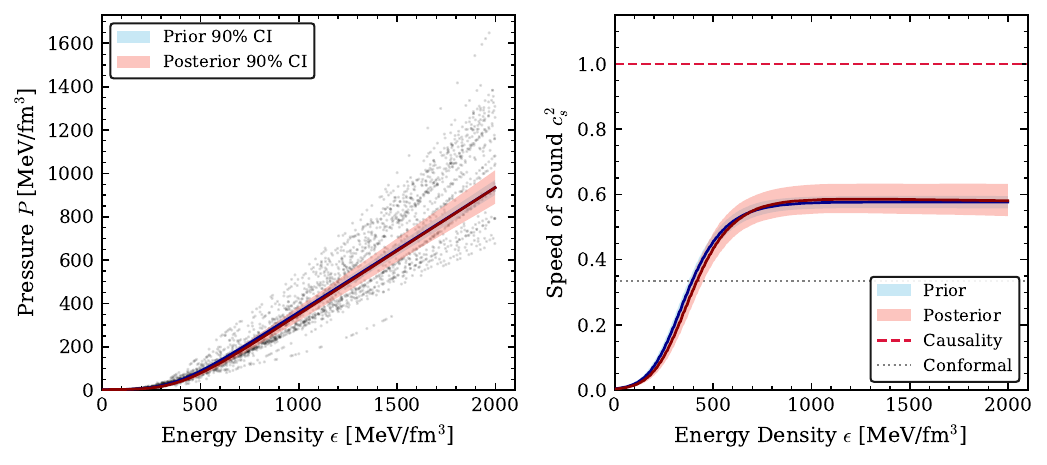}
    \caption{Microphysical predictions of the PI-BNN posterior ensemble. Left: pressure as a function of energy density, where the solid blue curve marks the posterior mean and the shaded blue band denotes the nominal 90\% variational credible interval. Right: the corresponding squared speed of sound $c_s^2$, where the solid blue curve gives the mean prediction and the red dashed line marks the causality limit $c_s^2=1$.}
    \label{fig:eos_posterior}
\end{figure*}

The primary objective of the Physics-Informed Bayesian Neural Network (PI-BNN) is to map the epistemic uncertainty of the dense-matter Equation of State (EoS) while enforcing fundamental physical limits. Figure \ref{fig:eos_posterior} presents the microscopic results of the PI-BNN posterior ensemble. 

In the left panel, the reconstructed pressure $P(\epsilon)$ is shown alongside the nominal 90\% variational credible interval. At lower densities, the network predictions are tightly constrained by the hadronic training ensemble and the three low-density nuclear anchors. They are also qualitatively compatible with low-density chiral-EFT guidance, although that comparison remains indirect because the microscopic benchmarks are usually formulated for pure neutron matter or symmetric nuclear matter rather than charge-neutral, beta-equilibrated neutron-star matter \cite{oertel/2017, chatziioannou/2025}. As the energy density increases into the ultra-dense core regime ($\epsilon \gtrsim 1000 \text{ MeV/fm}^3$), the variational band naturally widens. This broadening reflects increasing epistemic uncertainty in regions with sparse training data, where the inference is guided primarily by perturbative QCD (pQCD) boundary information.

The right panel of Figure \ref{fig:eos_posterior} shows the squared speed of sound, $c_s^2 = (\partial P / \partial \epsilon)_S$, which is the direct network output in the present formulation, plotted against the energy density. The mean $c_s^2$ rises steeply from the crust-core region, crosses the conformal value $c_s^2=1/3$ near $\epsilon \approx 500\,\text{MeV/fm}^3$, and levels off at a plateau of $c_s^2 \approx 0.58$-$0.61$ above $\epsilon \approx 1000\,\text{MeV/fm}^3$. The posterior shows a slight, statistically mild decline of the mean above $\epsilon \approx 1500\,\text{MeV/fm}^3$; we note that the parametrization does not forbid such a non-monotonic $c_s^2$, since only the pressure is constrained to be monotonic. By construction $c_s^2$ remains below the causality limit $c_s^2 = 1.0$ for every posterior draw. Physically, the plateau near $c_s^2\approx 0.6$ indicates moderately stiff matter, well above the conformal value, as required to support massive stars while remaining causal.

This density dependence admits a direct physical interpretation. Near and somewhat above saturation density, the posterior remains comparatively soft, consistent with the hadronic training ensemble and the radii inferred for canonical-mass stars. Above $\epsilon \approx 500\,\text{MeV/fm}^3$ the speed of sound rises past the conformal value and settles onto a plateau near $c_s^2\approx 0.6$, corresponding to the additional pressure support needed to sustain stars near the observed $2\,M_\odot$ threshold. At the highest densities reached in the posterior, the mean $c_s^2$ shows a mild decline rather than indefinite stiffening, indicating that the inferred EoS balances massive-star support with the high-density guidance from the pQCD anchor and causality.

Relative to the conformal value $c_s^2=1/3$, the posterior does not remain locked to conformal behavior across the full density range. Instead, it rises above $1/3$ at intermediate densities, signaling moderate stiffening above saturation density, and then bends downward at the highest densities reached by the ensemble, indicating partial softening while remaining comfortably below the causal bound.

\subsection{The Mass-Radius and Tidal-Deformability Posterior}
\label{sec:results_mr}
\begin{figure*}[t]
    \centering
    \includegraphics[width=\textwidth]{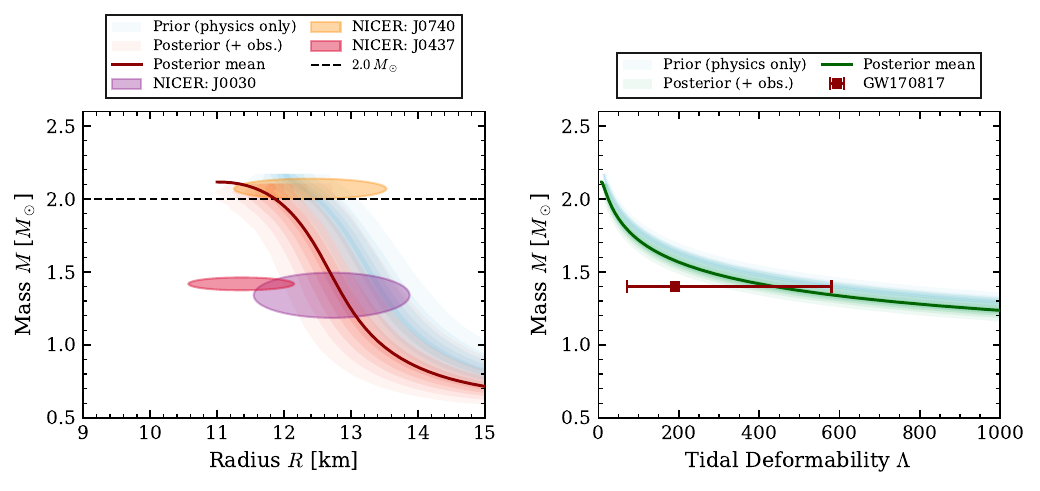}
    \caption{Joint observable-plane predictions from the PI-BNN, showing the physics-only prior (blue) and the observationally updated posterior (salmon/green) as nested nominal 90\% variational credible bands, each truncated at the median maximum mass to avoid survivorship bias. Left: the mass-radius ($M$-$R$) plane, with the $2.0\,M_\odot$ line and the NICER measurements for PSR J0030+0451, PSR J0740+6620, and PSR J0437-4715. Right: the mass-tidal-deformability ($M$-$\Lambda$) plane, with the GW170817 constraint. The contraction and shift of the posterior band relative to the prior visualizes the information added by the multi-messenger data: the canonical radius moves from $R_{1.4}=13.41$ to $12.74^{+0.97}_{-0.73}\,\text{km}$ and the tidal deformability from $\Lambda_{1.4}\simeq 500$ to $428^{+249}_{-130}$ (nominal 90\% variational CI), while the maximum mass remains above $2\,M_\odot$.}
    \label{fig:mr_relation}
\end{figure*}

By integrating the coupled TOV-plus-tidal equations over the ensemble of causally stable EoS samples, we project the microphysical uncertainties into the observable planes shown in Figure \ref{fig:mr_relation}. The left panel displays the mass-radius ($M$-$R$) relation, while the right panel shows the corresponding mass-tidal-deformability ($M$-$\Lambda$) relation. To avoid survivorship bias, the nominal 90\% variational credible bands are strictly truncated at the median maximum mass of the posterior, preventing statistical artifacts associated with the collapse of softer EoSs at lower masses.

The low-mass regime ($M \lesssim 0.5 M_\odot$) exhibits a sharp expansion in radius, characteristic of stellar structures dominated by the nuclear envelope. This behavior is consistent with the adopted crust-core stitching procedure and indicates that low-mass configurations are not artificially truncated.

The mass-radius band remains well behaved at low masses, with no anomalous tilt toward small radii: the smooth, strictly monotonic $c_s^2$-integral EoS yields a regular low-mass branch across the ensemble.

We first describe the physics-only (prior) posterior, and then, in Sec.~\ref{sec:reweighting}, its observational update; Fig.~\ref{fig:mr_relation} shows both bands overlaid in each plane.

Mass-Radius Plane: The left panel shows that the physics-only PI-BNN posterior extends above the observed $2.0\,M_\odot$ lower bound without any ad hoc high-density stiffening. The nominal 90\% variational credible interval overlaps with the NICER credible regions for PSR J0030+0451, PSR J0740+6620, and PSR J0437-4715, which are shown for comparison. The physics-only band gives a canonical radius $R_{1.4}=13.41\,\text{km}$ and a maximum mass $M_{\rm max}\gtrsim 2.0\,M_\odot$.

Tidal-Deformability Plane: The right panel shows that tidal deformability is a central posterior observable. The posterior displays the expected decrease of $\Lambda$ with increasing mass. The physics-only band gives $\Lambda_{1.4}\simeq 500$; the GW170817 reference value is $\Lambda_{1.4}=190^{+390}_{-120}$ (90\% CI). The prior value reflects the familiar tension between the low GW170817 deformability and the stiffness required to support $2\,M_\odot$ pulsars; the observational update in Sec.~\ref{sec:reweighting} lowers the posterior deformability toward the gravitational-wave constraint.

\subsection{Observational Inference of the Equation of State}
\label{sec:reweighting}

The construction described so far defines a physics-informed prior over EoS functions. We now perform a genuine observational inference by including current multi-messenger data directly in the variational objective, so that the network weights — and hence the posterior EoS itself — are updated by the data rather than merely reweighted after the fact.

The obstacle to placing an astrophysical likelihood inside the ELBO is that the map from an EoS to its observables $(M,R,\Lambda)$ requires integrating the TOV-plus-tidal system, which is not differentiable through the standard adaptive solver. We overcome this with a lightweight differentiable emulator: a small deterministic multilayer perceptron that maps a feature vector of the EoS — the pressures at a fixed set of reference densities — to the mass-radius and mass-tidal-deformability curves and the maximum mass. The emulator is trained once on the physics-only posterior ensemble, augmented with uniformly rescaled variants of each sampled EoS so that it learns the correct, monotonic dependence of the observables on the overall stiffness across the range explored during inference. It reproduces the results of exact solver within $0.15\,\text{km}$ at $R_{1.4}$ and is fully differentiable, so the observational log-likelihood can be added as a \texttt{pyro.factor} term evaluated on the emulated observables during SVI.

The observational likelihood combines the NICER mass-radius measurements of PSR J0030+0451, PSR J0740+6620, and PSR J0437-4715, the GW170817 tidal-deformability constraint at $1.4\,M_\odot$, and a one-sided penalty enforcing the $2\,M_\odot$ maximum-mass bound. To keep the inference stable, we warm-start it from the converged physics-only posterior and ramp the observational weight from zero to one over the first few hundred steps. All observables reported below are recomputed with the \emph{exact} TOV-plus-tidal solver on the resulting posterior ensemble; the emulator is used only to make the training likelihood differentiable.

Table~\ref{tab:obs_update} summarizes the effect of the observational update. The data pull the canonical-mass radius from $R_{1.4}=13.41\,\text{km}$ in the physics-only prior to $R_{1.4}=12.74^{+0.97}_{-0.73}\,\text{km}$ (nominal 90\% variational CI) in the observationally updated posterior, and give $\Lambda_{1.4}=428^{+249}_{-130}$, while the maximum mass remains above $2\,M_\odot$. The updated posterior remains within the dispersion of the hadronic training ensemble at all densities, so the softening it represents is data-driven rather than an extrapolation beyond the models. As an independent check, importance-reweighting the prior ensemble by the same likelihood yields a consistent canonical radius ($R_{1.4}\simeq 12.9\,\text{km}$), and a leave-one-out analysis shows that no single measurement dominates the result, the largest individual effect being the removal of PSR J0437-4715 ($+0.2\,\text{km}$). The shift between the prior and posterior bands, shown in both observable planes in Fig.~\ref{fig:mr_relation}, is a direct measure of the information content of the current multi-messenger dataset. Because the emulator makes the likelihood differentiable and cheap to evaluate, this update requires no re-solution of the stellar-structure equations during training and can be repeated, or extended with additional likelihoods, as measurements improve.

We assessed the calibration of the variational posterior through a held-out coverage test, retraining on 80\% of the hadronic tables and checking how often the resulting bands contain the withheld EoSs. The nominal 90\% variational interval contains only $\approx15\%$ of the held-out EoSs over the density range constrained by the data, demonstrating substantial underdispersion. The probability-integral transform is centred (mean $0.50$), which indicates no detectable systematic rank shift toward softer or stiffer EoSs in this test, but does not establish that the point estimates are unbiased. We therefore label the reported PI-BNN bands as \emph{nominal} variational credible intervals and interpret their widths as lower bounds on the uncertainty represented by the present mean-field approximation. Richer variational families or sampling-based inference, including Hamiltonian Monte Carlo, may reduce this underdispersion but require additional computational and convergence studies.

\begin{table}[t]
\centering

\begin{tabular}{lccc}
\hline\hline
 & $R_{1.4}$ [km] & $\Lambda_{1.4}$ & $M_{\rm max}$ [$M_\odot$] \\
\hline
Prior (physics only) & $13.41$ & $\simeq 500$ & $\gtrsim 2.0$ \\
Posterior ($+$ obs.) & $12.74^{+0.97}_{-0.73}$ & $428^{+249}_{-130}$ & $\gtrsim 2.0$ \\
\hline\hline
\end{tabular}

\caption{Canonical-mass observables before and after the observational update, computed with the exact TOV-plus-tidal solver on the posterior ensemble. Quoted uncertainties are nominal 90\% variational credible intervals.}
\label{tab:obs_update}
\end{table}

\section{Conclusions}
\label{sec:conclusions}

In this work, we introduced a physics-informed Bayesian Neural Network (PI-BNN) to infer the dense matter equation of state and systematically propagate microscopic uncertainties into macroscopic mass-radius and tidal-deformability observables. By treating the EoS as a continuous probability distribution over functions rather than a rigid parametric curve, the method reduces derivative artifacts and structural bias associated with traditional piecewise-polytropic and spectral representations while retaining fully Bayesian function-space inference.

By parametrizing the EoS through its speed of sound, $c_s^2(\epsilon)=\mathrm{sigmoid}[g_\theta(\epsilon)]$, and obtaining the pressure by integration, the method guarantees causality, thermodynamic stability, positivity, and monotonicity for every core draw. This eliminates monotonicity and causality penalties as well as EoS-level post-sampling rejection: every posterior core draw is retained. Anchor terms at three low-density nuclear points and at the perturbative-QCD boundary fix the normalization of the EoS, and the BNN-to-SLy4 stitching yields stable TOV solutions across both low-mass and high-mass stars.

Beyond constructing this physics-informed prior, we performed a genuine observational inference by embedding the multi-messenger likelihood — the NICER mass-radius measurements, the GW170817 tidal-deformability constraint, and the $2\,M_\odot$ maximum-mass bound — directly in the variational objective. A differentiable emulator of the TOV-plus-tidal map, validated against the exact solver to within $0.15\,\text{km}$ in $R_{1.4}$, makes this update tractable inside the ELBO. The data move the canonical-mass radius from $R_{1.4}=13.41$ to $12.74^{+0.97}_{-0.73}\,\text{km}$ and the tidal deformability from $\Lambda_{1.4}\simeq 500$ to $428^{+249}_{-130}$ (nominal 90\% variational CI), while the maximum mass remains above $2\,M_\odot$; the posterior thus overlaps the NICER credible regions, including the refined J0437-4715 measurement, and is consistent with the GW170817 constraint. An independent importance-reweighting analysis and a leave-one-out test confirm that the central result is robust and not driven by any single measurement. The shift between the prior and posterior bands quantifies the information content of the current data, while the held-out coverage test shows that their widths remain underdispersed under the present mean-field approximation. The framework provides a scalable, non-parametric route for incorporating future multi-messenger constraints directly into EoS inference, so that improved X-ray, radio, and gravitational-wave measurements can be folded in transparently as they become available.

A limitation of the present analysis is that the training set is restricted to hadronic EoSs; explicit phase transitions and hybrid-matter EoSs are not included. This scope was chosen to establish a controlled baseline with a uniform microphysical dataset and to validate end-to-end uncertainty propagation before introducing additional transition parameters. Extending the framework to include phase transitions and hybrid-matter families is a natural next step and will require dedicated priors and expanded training data.
\bigskip

{\bf Acknowledgments}

This work was supported by U.S. Department of Energy Office of Nuclear Physics under Contract No. DE-SC0026074 with East Texas A\&M University. RVL was supported by INCT-FNA (Instituto Nacional de Ci\^ encia e Tecnologia, F\' isica Nuclear e Aplica\c c\~oes), research Project No. 464898/2014-5, and also thanks to CAPES/FAPERJ/CNPq.

\bibliography{bibliography}

\end{document}